\documentstyle[preprint,tighten,aps,epsfig]{revtex}

\begin{document}
\title{Microscopic $NN\rightarrow NN^{\ast }(1440)$ transition
potential: Determination of $\pi NN^{\ast}(1440)$ and $\sigma
NN^{\ast}(1440) $ coupling constants.}
\author{B. Juli\'a-D\'{\i}az $^{1}$, A. Valcarce $^{1,2}$, P. Gonz\'alez$^{2}$, and
F. Fern\'andez $^{1}$}
\address{$^1$ Grupo de F\' \i sica Nuclear \\
Universidad de Salamanca, E-37008 Salamanca, Spain}
\address{$^2$ Dpto. de F\' \i sica Te\'orica and IFIC\\
Universidad de Valencia - CSIC, E-46100 Burjassot, Valencia, Spain}
\maketitle

\begin{abstract}
A $NN\rightarrow NN^{\ast }(1440)$ transition potential, based on an
effective quark-quark interaction and a constituent quark cluster model for
baryons, is derived in the Born-Oppenheimer approach. The potential shows
significant differences with respect to those obtained by a direct scaling
of the nucleon-nucleon interaction. From its asymptotic behavior we extract
the values of $\pi NN^{\ast }(1440)$ and $\sigma NN^{\ast }(1440)$ coupling
constants in a particular coupling scheme.
\end{abstract}

\vspace*{2cm} \noindent Keywords: \newline
nonrelativistic quark models, baryon-baryon potentials \newline
\newline
\noindent Pacs: \newline
12.39.Jh, 13.75.Cs, 14.20.Gk, 24.85.+p

\narrowtext
\newpage

\section{Introduction}

The nucleon-nucleon ($NN$) interaction constitutes the basic process in
nuclear dynamics and as such it has been for many years the object of
extensive study. Different approaches, going from almost completely
phenomenological potentials and meson-exchange treatments at the baryon
level to quark model descriptions, have been developed to mitigate the
current impossibility to directly obtain the form of the interaction from
QCD. Each approach has its own justification. The use of more and more
sophisticated phenomenological baryonic potentials allows a very precise fit
of some data in a selected energy domain. Meson-exchange approaches at the
baryon level make clear the role of effective hadronic degrees of freedom at
a given energy scale. Quark model descriptions based on QCD and formulated
in terms of effective quark degrees of freedom might be the closest approach
to the underlying theory.

From all of them we have been able to reach a quite reasonable, though not
complete, understanding of the two-nucleon interaction at low energy ($%
T_{lab}\leq 300$ MeV ) \cite{PARI,BONN,NIJM,SAL1,SAL2}. When increasing the
energy, the opening of channels involving the excitation of baryon
resonances determines to a good extent the character of the interaction. Up
to 1 GeV relative kinetic energy in the laboratory, the $\Delta (1232)$ and $%
N^{\ast }(1440)$ are the most prominent resonances \cite{WEAR}. The role
played by the $\Delta $ resonance has been studied at the baryon level \cite
{SUHI,AREN,GREE,HOMA,LOMO,HLEE,WISA,PEHS,HAHJ} as well as at the quark level 
\cite{PLB1,SAUE,HUM1}, by means of $NN\rightarrow N\Delta $, $N\Delta
\rightarrow N\Delta $, and $\Delta \Delta \rightarrow \Delta \Delta $
potentials. These studies show the relevance of a quark analysis to properly
treat the short-range part of the interaction. 

The $N^{\ast }(1440)$ (Roper)
is a broad resonance which couples strongly (60$-$70$\%$) to the $\pi N$
channel and significantly (5$-$10$\%$) to the $\sigma N$ channel \cite{PDGB}.
These features suggest that the Roper resonance should play an important
role in nuclear dynamics as an intermediate state. This role has been
analyzed at the baryon level. Graphs involving the excitation of $N^{\ast
}(1440)$ appear in different systems, as for example the neutral pion
production in proton-proton reactions \cite{PENA} or the three-nucleon
interaction mediated by $\pi $ and $\sigma $ exchange contributing to the
triton binding energy \cite{COON}. The excitation of the Roper resonance has
also been used to explain the missing energy spectra in the $%
p(\alpha,\alpha^{\prime})$ reaction \cite{OSET} or the $np\to d(\pi\pi)^0$
reaction \cite{RUSO}. The coupling of the $N^{\ast }(1440)$ to $\pi N$ and $%
\sigma N$ channels could also be important in heavy ion collisions at
relativistic energies \cite{LIKO,HUBER}. The presence of $NN^{\ast }(1440)$
configurations on the deuteron has been suggested long ago \cite
{HAJO,REID,ROST,ARDW}. Finally, pion electro- and photoproduction may take
place through the $N^*(1440)$ excitation \cite{GARCI}. However the use of a $%
NN\rightarrow NN^{\ast }(1440)$ transition potential as a straightforward
generalization of some pieces of the $NN\rightarrow NN$ potential plus the
incorporation of resonance width effects may have, as commented above for 
the $\Delta $, serious shortcomings specially concerning the
short-range part of the interaction \cite{PLB1,SAUE,HUM1,BJUL}.

In view of the current interest in nucleon resonances in a nuclear physics
context, it seems appropriate to extend the quark model $NN$ calculations to
treat all presently accepted $N^*$ resonances. In this article we propose a
quark model treatment of the $NN\rightarrow NN^{\ast }(1440)$ interaction.
We shall adopt the same quark model approach previously used for the $\Delta$
case and also applied to the $NN^*(1440) \to NN^*(1440)$ interaction \cite
{BJUL}. We shall center our attention in the derivation of a $NN \to
NN^*(1440)$ transition potential from a quark-quark ($qq$) basic interaction
incorporating gluon, pion and sigma exchanges. For the sake of simplicity we
shall follow a Born-Oppenheimer (BO) approximation with harmonic oscillator
baryon wave functions written in terms of quarks. The Roper resonance, $%
N^{\ast }(1440)$, will be considered as a stable particle.

A main feature of our quark treatment is its universality in the sense that
all the baryon-baryon interactions are treated on an equal footing.
Moreover, once the model parameters are fixed from $NN$ data
there are no free parameters for any other case. This allows a
microscopic understanding and connection of the different baryon-baryon
interactions that is beyond the scope of any analysis based only on
effective hadronic degrees of freedom. This is important not only in the
short-range regime, where it does not exist a definite prescription for the
potentials at the baryon level when resonances are involved, but at all
distances. In particular, the asymptotic (long-range) behavior of the $NN
\to NN^*(1440)$ potential allows the determination 
of the $\pi NN^{\ast }(1440)$ and $\sigma NN^{\ast }(1440)$ coupling
constants as well as their ratios to the $\pi NN$ and $%
\sigma NN$ coupling constants, respectively. These studies are instructive
inasmuch as they are expected to lead to a deeper understanding of the
nuclear potential and entail a rethinking of basic nuclear concepts from the
point of view of the fundamental quark substructure.

The article is organized as follows. In Sec. II we write the 
$qq$ interaction and analyze the two-baryon wave functions in order to
obtain the $NN\rightarrow NN^{\ast }(1440)$ transition potential. In Sec.
III we draw the results for different partial waves and spin-isospin
channels. In Sec. IV we proceed to determine the $\pi NN^{\ast }(1440)$ and $%
\sigma NN^{\ast}(1440)$ coupling constants and relate them to the $\pi NN$
and $\sigma NN$ coupling constants. Finally, in Sec. V we summarize our main
conclusions.

\section{$NN\rightarrow NN^{\ast }(1440)$ transition potential}

In the Born-Oppenheimer approximation the $NN\rightarrow NN^{\ast}(1440)$
transition potential at the interbaryon distance $R$ is obtained by
sandwiching between $NN$ and $NN^{\ast }(1440)$ states (expressed in terms
of quarks) the $qq$ potential for all the pairs formed by two quarks
belonging to different baryons. In other words:

\begin{equation}
V_{N N (L \, S \, T) \rightarrow N N^* (L^{\prime}\, S^{\prime}\, T)} (R) =
\xi_{L \,S \, T}^{L^{\prime}\, S^{\prime}\, T} (R) \, - \, \xi_{L \,S \,
T}^{L^{\prime}\, S^{\prime}\, T} (\infty) \, ,  \label{Pot}
\end{equation}

\noindent where

\begin{equation}
\xi_{L \, S \, T}^{L^{\prime}\, S^{\prime}\, T} (R) \, = \, {\frac{{\left
\langle \Psi_{N N^*}^{L^{\prime}\, S^{\prime}\, T} ({\vec R}) \mid
\sum_{i<j=1}^{6} V_{qq}({\vec r}_{ij}) \mid \Psi_{N N}^{L \, S \, T} ({\vec R%
}) \right \rangle} }{{\sqrt{\left \langle \Psi_{N N^*}^{L^{\prime}\,
S^{\prime}\, T} ({\vec R}) \mid \Psi_{N N^*}^{L^{\prime}\, S^{\prime}\, T} ({%
\vec R}) \right \rangle} \sqrt{\left \langle \Psi_{N N}^{L \, S \, T} ({\vec %
R}) \mid \Psi_{N N}^{L \, S \, T} ({\vec R}) \right \rangle}}}} \, .
\end{equation}

The quark-quark potential has been very much detailed elsewhere \cite
{SAL1,SAL2} and it will only be written here for completeness. It reads,

\begin{equation}
V_{qq}(\vec{r}_{ij})=V_{CON}(\vec{r}_{ij})+V_{OGE}(\vec{r}_{ij})+V_{OPE}(%
\vec{r}_{ij})+V_{OSE}(\vec{r}_{ij}) \,,  \label{terms}
\end{equation}

\noindent where $\vec{r}_{ij}$ is the interquark distance. $V_{CON}$ is the
confining potential, whose detailed radial structure being fundamental to
study the hadron spectra is expected to play a minor role for the two-baryon
interaction \cite{SHIM}. To be consistent with baryon and meson spectroscopy
it will be taken to be linear

\begin{equation}
V_{CON}(\vec{r}_{ij})=-a_{c}\,\vec{\lambda _{i}}\cdot \vec{\lambda _{j}}%
\,r_{ij}\, ,
\end{equation}

\noindent where the $\lambda ^{\prime }s$ stand for the color SU(3)
matrices. $V_{OGE}$ is the perturbative one-gluon-exchange (OGE) interaction
containing Coulomb ($\frac{1 }{r_{ij}}$), spin-spin (${\vec{\sigma}}%
_{i}\cdot {\vec{\sigma}}_{j})$ and tensor terms ($S_{ij}$) \cite{RUJU}

\begin{equation}
V_{OGE}({\vec{r}}_{ij})={\frac{1}{4}}\,\alpha _{s}\,{\vec{\lambda}}_{i}\cdot 
{\vec{\lambda}}_{j}\Biggl \lbrace{\frac{1}{r_{ij}}}-{\frac{\pi }{m_{q}^{2}}}%
\,\biggl [1+{\frac{2}{3}\vec{\sigma}}_{i}\cdot {\vec{\sigma}}_{j}\biggr ]%
\,\delta ({\vec{r}}_{ij})-{\frac{3}{{4m_{q}^{2}\,r_{ij}^{3}}}}\,S_{ij}\Biggr
\rbrace\,,
\end{equation}

\noindent and $V_{OPE}$ and $V_{OSE}$ are the one-pion (OPE) and one-sigma
exchange (OSE) interactions given by:

\begin{eqnarray}
V_{OPE} ({\vec r}_{ij}) & = & {\frac{1 }{3}} \, \alpha_{ch} {\frac{\Lambda^2 
}{\Lambda^2 - m_\pi^2}} \, m_\pi \, \Biggr\{ \left[ \, Y (m_\pi \, r_{ij}) - 
{\frac{ \Lambda^3 }{m_{\pi}^3}} \, Y (\Lambda \, r_{ij}) \right] {\vec \sigma%
}_i \cdot {\vec \sigma}_j +  \nonumber \\
& & \left[ H( m_\pi \, r_{ij}) - {\frac{ \Lambda^3 }{m_\pi^3}} \, H( \Lambda
\, r_{ij}) \right] S_{ij} \Biggr\} \, {\vec \tau}_i \cdot {\vec \tau}_j \, ,
\end{eqnarray}

\begin{equation}
V_{OSE}({\vec{r}}_{ij})=-\alpha _{ch}\,{\frac{4\,m_{q}^{2}}{m_{\pi }^{2}}}{%
\frac{\Lambda ^{2}}{\Lambda ^{2}-m_{\sigma }^{2}}}\,m_{\sigma }\,\left[
Y(m_{\sigma }\,r_{ij})-{\frac{\Lambda }{{m_{\sigma }}}}\,Y(\Lambda \,r_{ij})%
\right] \,,
\end{equation}

\noindent where $\Lambda $ is a cutoff parameter and

\begin{equation}
Y(x)=\frac{e^{-x}}{x} \, ,
\end{equation}

\begin{equation}
H(x)=\left( 1+\frac{3}{x}+\frac{3}{x^{2}}\right) \,Y(x) \, .
\end{equation}

The values chosen for the parameters are tabulated in Table \ref{table1}.
They are taken from Ref. \cite{SAL2} where an accurate
description of the $NN$ scattering phase shifts and the deuteron properties
is obtained. They also provide a
reasonable description of the baryon spectrum \cite{HUM3}.

The Born-Oppenheimer approximation followed integrates out the quark
coordinates keeping R fixed. Hence, quantum fluctuations of the two-baryon
center-of-mass are neglected. Nonetheless, a more complete treatment as the
one implied by the use of the resonating group method may not represent, at
least for the calculations we perform, major changes as it turns out to be
the case for the $NN$ interaction \cite{HUM2}.

The $NN$ and $NN^{\ast }(1440)$ wave functions we shall use hereforth have
been also detailed elsewhere \cite{BJUL}. Here we only quote some results
that will be useful in what follows. The $N$ and $N^{\ast }(1440)$ states
are given in terms of quarks by:

\begin{equation}
|N\rangle = |[3](0s)^3\rangle \otimes [1^{3}]_{c} \, ,
\label{eq10}
\end{equation}

\begin{equation}
|N^*(1440)\rangle = \left\{ \sqrt{\frac{2}{3}}|[3](0s)^2(1s)\rangle -\sqrt{%
\frac{1}{3}} |[3](0s)(0p)^2\rangle \right\} \otimes [1^{3}]_{c} \, ,
\label{eq11}
\end{equation}

\noindent where $[1^3]_{c}$ is the completely antisymmetric color state, $%
[3] $ is the completely symmetric spin-isospin state and $0s$, $1s$, and $0p$%
, stand for harmonic oscillator orbitals.

For a definite orbital angular momentum $L$ and spin $S$, the $NN$ wave
function satisfies, due to the identity of the baryons, the selection rule

\begin{equation}
L_{NN}-S_{NN}-T_{NN}= {\rm odd} \, .
\end{equation}

This is not the case for the $NN^{\ast }(1440)$ system, due to the
nonidentity of $N$ and $N^{\ast }(1440)$. Nevertheless antisymmetry at quark
level, coming from the identity of quarks, gives rise to a generalized
selection rule for any nucleon resonance $N^*$, that can be written as

\begin{equation}
L_{NN^{\ast }}-S_{NN^{\ast }}-T_{NN^{\ast }}+f= {\rm odd} \, ,
\end{equation}

\noindent where $f$ is the $NN^{\ast }$ spin-isospin parity determining the
symmetric ($f=$ even) or antisymmetric ($f=$ odd) character of the $NN^{\ast
}$ wave function in the spin-isospin space. The case $f=$ even gives rise to
the same $NN^{\ast }$ channels than in the $NN$ case, whereas the case $f=$
odd corresponds to channels forbidden in the $NN$ case that reflects the
effects of quark identity beyond baryon identity. These {\it forbidden}
channels play a relevant role in the $NN^\ast (1440)\rightarrow NN^\ast
(1440)$ case \cite{BJUL}. However for the $NN\rightarrow NN^{\ast}(1440)$
transition we are dealing with the situation simplifies considerably. In
fact, as the strong interaction preserves isospin we have $T_{NN^{\ast
}}=T_{NN}$. Furthermore the structure of the interaction given by Eq. (\ref
{terms}) allows only to connect $NN$ and $NN^{\ast}(1440)$ channels
verifying $L^{\prime}- L = 0$ or $2= S^{\prime}-S$. Therefore the initial
state selection rule translates to the final state, i.e. only $f=$ even $%
NN^{\ast}(1440)$ channels are allowed.

The most representative diagrams contributing to the $NN\rightarrow NN^{\ast
}(1440)$ potential, as calculated from Eq. (\ref{Pot}), are drawn in Fig. 
\ref{fig1}. We distinguish between the direct diagrams (labeled as $V_{36}$
in Fig. \ref{fig1}), not involving quark exchanges, and the rest of diagrams
including exchange of quarks (labeled as $V_{ij}P_{36}$ in Fig. \ref{fig1}).
Most diagrams contributing to the interaction are due to the first term of
the $N^*(1440)$ wave function ($|[3](0s)^2(1s)\rangle$), only a few of them,
those with two vertical dashed lines, correspond to the second term of the $%
N^*(1440)$ wave function ($|[3](0s)(0p)^2\rangle$).

\section{Results}

In Figs. \ref{1s0de}, \ref{1p1de}, and \ref{1d2de}, we show the potentials
obtained for $L=0$ ($^{1}S_{0}$ and $^{3}S_{1}$), $L=1$ ($^{1}P_{1}$ and $%
^{3}P_{0}$), and $L=2$ ($^{1}D_{2}$ and $^{3}D_{1}$) partial waves,
respectively. Contributions from the different terms of the potential as
separated in Eq. (\ref{terms}) have been made explicit. For some selected
partial waves, we separate in Fig. \ref{3s1di} the contribution of the
different diagrams depicted in Fig. \ref{fig1}. Let us mention that an
arbitrary global phase between the $N$ and $N^*(1440)$ wave functions 
as written in Eqs. (\ref{eq10}) and (\ref{eq11}) has to be
chosen. We will discuss all aspects depending on this choice.

There are a number of general features that can be enumerated:

(i) The very long-range part of the interaction ($R>4$ fm ) comes dominated,
as for the $NN\rightarrow NN$ and $NN^{\ast }(1440)\rightarrow NN^{\ast
}(1440)$ cases, by the one-pion exchange, the longest-range piece of the
potential. However the asymptotic potential reverses sign with respect to
both $NN\rightarrow NN$ and $NN^{\ast }(1440)\rightarrow NN^{\ast }(1440)$.
Thus for $S$ and $D$ waves the $NN\rightarrow NN^{\ast }(1440)$ interaction
is asymptotically repulsive. This sign reversal is a direct consequence of
the presence of a node in the $N^{\ast }(1440)$ wave function what implies a
change of sign with respect to the $N$ wave function at long distances (if
the opposite sign for the $N^*(1440)$ wave function were chosen the very
long-range part of the interaction would be attractive but there would also
be a change in the character of the short-range part). This is also
corroborated by the study of the one-sigma exchange interaction that is
always asymptotically repulsive at difference to the $NN\rightarrow NN$ and $%
NN^*(1440)\to NN^*(1440)$ cases (for $NN^*(1440)\to NN^*(1440)$ there are
two compensating changes of sign coming from the two Ropers).

It is worth to remark that no quark antisymmetrization effects survive
either in the numerator or in the denominator (norm) of Eq. (\ref{Pot}) at
these distances. In other words, the potential corresponds to a direct
baryon-baryon interaction.

(ii) For the long-range $2<R<4$ fm part, the one-pion and one-sigma-exchange
potentials altogether determine the character of the interaction, since the
one-gluon-exchange gives a negligible contribution for $R\geq 2$ fm. One
should also notice that although the contribution from quark exchange
diagrams is very much suppressed for $R\geq 2$ fm, some quark
antisymmetrization effects may still be present through the norm (see Fig. 1
of Ref. \cite{BJUL}).

(iii) At intermediate range $0.6<R<2$ fm a complex interplay among all
pieces of the potential (gluon, pion and sigma) generates the final form of
the interaction. When decreasing $R$ from 2 fm to 0.6 fm two effects take
place. On the one hand, quark exchange diagrams are increasingly important
becoming dominant below $R=1.5$ fm. On the other hand the different pieces
of the potential are changing sign: from attractive to repulsive for the
gluon in all partial waves, from repulsion to attraction for the sigma in $S$
and $D$ waves and from repulsion to attraction and again to repulsion for
the pion in $S$ and $D$ waves. As a combined result of these effects the
total potential turns out to be attractive from $R=1.5$ fm down to a lower
value of $R$ different for each partial wave. This behavior, related again
to the node in the Roper wave function, contrasts with the $NN\rightarrow NN$
and $NN^{\ast}(1440)\rightarrow NN^{\ast}(1440)$ cases, where for instance
for $S$ and $D$ waves the scalar (sigma) part keeps always the same sign and
gives the dominant contribution for $R>0.8$ fm.

(iv) The choice of $0.6$ fm as a lower limit for the intermediate range
comes motivated by the repulsive character of the potential in all partial
waves for shorter distances. The one-gluon and one-pion quark exchange parts
are mainly responsible for such a repulsion as it turns out to be the case
for $NN\rightarrow NN$ and $NN^{\ast}(1440)\rightarrow NN^{\ast}(1440)$.
Nevertheless there are two distinctive features with respect to these cases:
in $NN\rightarrow NN^{\ast}(1440)$ the intensity of the repulsion at $R=0$
and the value of $R$ at which the interaction becomes repulsive are
significantly lower than in $NN\rightarrow NN$ and $NN^{\ast}(1440)%
\rightarrow NN^{\ast}(1440)$. This is a clear effect of the more similarity
(higher overlap) in these cases between initial and final states what makes
the Pauli principle more active.

\section{$\protect\pi NN^{\ast}(1440)$ and $\protect\sigma NN^{\ast}(1440)$
coupling constants}

The potential obtained can be also written at all distances in terms of
baryonic degrees of freedom \cite{HOLI}. One should realize that a $qq$ spin
and isospin independent potential as for instance the scalar one-sigma
exchange, gives rise at the baryon level, apart from a spin-isospin
independent potential, to a spin-spin, an isospin-isospin and a spin-isospin
dependent interaction \cite{SAL1}. Nonetheless for distances $R\geq 4$ fm,
where quark antisymmetrization interbaryon effects vanish, we are only left
with the direct part, i.e. with a scalar one-sigma exchange at the baryon
level. The same kind of argument can be applied to the one-pion exchange
potential. Thus asymptotically ($R\geq 4$ fm) OSE and OPE have at the baryon
level the same spin-isospin structure than OSE and OPE at the quark level.
Hence we can parametrize the asymptotic central interactions as (the $%
\Lambda $ depending exponential term is negligible asymptotically as
compared to the Yukawa term) 
\begin{equation}
V_{NN\rightarrow NN^{\ast}(1440)}^{OPE}(R)=\frac{1}{3} \, \frac{g_{\pi NN}}{%
\sqrt{4\pi }} \, \frac{g_{\pi NN^{\ast}(1440)}}{\sqrt{4\pi }} \, \frac{%
m_{\pi }}{2M_{N}} \, \frac{m_{\pi }}{2(2M_{r})} \, \frac{\Lambda ^{2}}{%
\Lambda ^{2}-m_{\pi }^{2}} [(\vec{\sigma }_{N}.\vec{\sigma }_{N})(\vec{\tau }%
_{N}.\vec{\tau }_{N})] \, \frac{e^{-m_{\pi }R}}{R} \, ,  \label{lrg}
\end{equation}

\noindent and

\begin{equation}
V_{NN\rightarrow NN^{\ast}(1440)}^{OSE} (R)=- \, \frac{g_{\sigma NN}}{\sqrt{%
4\pi }} \, \frac{g_{\sigma NN^{\ast}(1440)}}{\sqrt{4\pi }} \, \frac{\Lambda
^{2}}{\Lambda ^{2}-m_{\sigma }^{2}} \, \frac{e^{-m_{\sigma }R}}{R} \, ,
\label{slrg}
\end{equation}

\noindent where $g_{i}$ stands for the coupling constants at the baryon
level and $M_{r}$ is the reduced mass of the $NN^{\ast }(1440)$ system $%
\left( \frac{1}{M_{r}}=\frac{1}{M_{N}}+\frac{1}{M_{N^{\ast }(1440)}}\right) $%
. One should note that at these distances the use of the BO
approximation is justified and the resonating group method potential would
give quite the same results.

By comparing these baryonic potentials with the asymptotic behavior of the
OPE and OSE previously obtained from the quark calculation we can extract
the $\pi NN^{\ast }(1440)$ and $\sigma NN^{\ast }(1440)$ coupling constants.
As the parameters at the quark level are fixed once for all from the $NN$
interaction our results allow a prediction of these constants in terms of
the elementary $\pi qq$ coupling constant and the one-baryon model dependent
structure. The sign obtained for the meson-$NN^{\ast }(1440)$ coupling
constants and for their ratios to the meson-$NN$ coupling constants 
is ambiguous since it comes determined by the arbitrarily chosen
relative sign between the $N$ and $N^{\ast }(1440)$ wave functions. Only
the ratios between the $\pi NN^{\ast }(1440)$ and $\sigma NN^{\ast }(1440)$
would be free of this uncertainty. This is why we will quote absolute values
except for these cases where the sign is a clear prediction of the model. To
get such a prediction we can use any partial wave. We shall use for
simplicity the $^{1}S_{0}$ wave, this is why we only wrote the central
interaction in Eq. (\ref{lrg}).

The ${\frac{ \Lambda ^{2}}{{\Lambda ^{2}-m_i^{2}}}}$ vertex factor comes
from the vertex form factor chosen at momentum space as a square root of
monopole $\left( \frac{\Lambda ^{2}}{\Lambda ^{2}+\vec{q}^{\, \, 2}} \right)
^{\frac{1}{2}}$, the same choice taken at the quark level, where chiral
symmetry requires the same form for pion and sigma. A different choice for
the form factor at the baryon level, regarding its functional form as well
as the value of $\Lambda $, would give rise to a different vertex factor and
eventually to a different functional form for the asymptotic behavior. For
instance, for a modified monopole form, $\left( \frac{\Lambda ^{2}-m^{2}}{%
\Lambda ^{2}-q^{2}}\right) ^{\frac{1}{2}}$, where $m$ is the meson mass ($%
m_{\pi }$ or $m_{\sigma }$), the vertex factor would be $1$, i.e. $\frac{%
\Lambda ^{2}-m^{2}}{\Lambda ^{2}-m^{2}}$, keeping the potential the same
exponentially decreasing asymptotic form. Then it is clear that the
extraction from any model of the meson-baryon-baryon coupling constants
depends on this choice. We shall say they depend on the coupling scheme.

For the one-pion exchange and for our value of $\Lambda =4.2$ fm$^{-1}$, $%
\frac{\Lambda ^{2}}{\Lambda ^{2}-m_{\pi }^{2}}=1.03$, pretty close to $1$.
As a consequence, in this case the use of our form factor or the modified
monopole form at baryonic level makes little difference in the determination
of the coupling constant. This fact is used when fixing 
$g_{\pi qq}^{2}/{4\pi }$ from the experimental value of 
$g_{\pi NN}^{2}/{4\pi }$ extracted from $NN$ data. 
The value we use for $\alpha _{ch}=\frac{m_{\pi }^{2}}{4m_{q}^{2}}
\frac{g_{\pi qq}^{2}}{4\pi }=\left( \frac{3}{5}\right) ^{2}%
\frac{g_{\pi NN}^{2}}{4\pi }\frac{m_{\pi }^{2}}{4m_{N}^{2}}
e^{-\frac{m_{\pi }^{2}b^{2}}{2}}=0.027$ 
corresponds to $g_{\pi NN}^{2}/{4\pi }=14.83$.

To get $\frac{g_{\pi NN^*(1440)}}{\sqrt{4\pi }}$ we turn to our numerical
results for the $^{1}S_{0}$ OPE potential, Fig. \ref{nr1s0lrg}(a), and fit
its asymptotic behavior (in the range $R:5\rightarrow 9$ fm) to Eq. (\ref
{lrg}). We obtain

\begin{equation}
\frac{g_{\pi NN}}{\sqrt{4\pi }} \frac{g_{\pi NN^{\ast}(1440)}}{\sqrt{4\pi }} 
\frac{\Lambda ^{2}}{\Lambda ^{2}-m_{\pi }^{2}}= \, - \, 3.73 \, ,
\end{equation}

\noindent i.e. $\frac{g_{\pi NN^{\ast }(1440)}}
{\sqrt{4\pi }}= - 0.94$. As
explained above only the absolute value of this coupling constant is well
defined. Let us note that in Ref. \cite{RIBO} a different sign with respect
to our coupling constant is obtained what is a direct consequence of
the different global sign chosen for the $N^{\ast }(1440)$ wave function.
The coupling scheme dependence can be explicitly eliminated if we compare $%
g_{\pi NN^{\ast }(1440)}$ with $g_{\pi NN}$ extracted from the $%
NN\rightarrow NN$ potential within the same quark model approximation, Fig. 
\ref{nr1s0lrg}(b). Thus we get

\begin{equation}
\left | \frac{g_{\pi NN^{\ast}(1440)}}{g_{\pi NN}} \right |=0.25 \,.
\label{eq17}
\end{equation}

By proceeding in the same way for the OSE potential, i.e. by fitting the
potential given in Fig. \ref{nr1s0silrg}(a) to Eq. (\ref{slrg}), and
following an analogous procedure for the $NN$ case, Fig. \ref{nr1s0silrg}%
(b), we can write

\begin{equation}
\left |\frac{g_{\sigma NN^{\ast}(1440)}}{g_{\sigma NN}} \right |=0.47 \, .
\label{eq18}
\end{equation}

The relative phase chosen for the $N^*(1440)$ wave function with respect
to the $N$ wave function is not
experimentally relevant in any two step process
comprising $N^{\ast }(1440)$ production and its subsequent decay. However it
will play a relevant role in those reactions where the same field ($\pi $ or 
$\sigma $) couples simultaneously 
to both systems, $NN$ and $NN^{\ast }(1440)$.
In these cases the interference term between both diagrams would
determine the magnitude of the cross section \cite{OSET}.

The ratio given in Eq. (\ref{eq17}) is similar to that obtained in Ref. \cite
{RIBO} and a factor 1.5 smaller than the one obtained from the analysis of
the partial decay width \cite{RIBO}. Nonetheless one can find in the
literature values for $f_{\pi NN^{\ast }(1440)}$ ranging between 0.27$-$0.47
coming from different experimental analyses with uncertainties
associated to the fitting of parameters \cite{RUSO,HUBER,GARCI}.

Regarding the ratio obtained in Eq. (\ref{eq18}), our result agrees quite
well with the only experimental available result, obtained in Ref. \cite
{OSET} from the fit of the cross section of the isoscalar Roper excitation
in $p(\alpha,\alpha^{\prime})$ in the 10$-$15 GeV region, where
a value of 0.48 is given.

Furthermore, we can give a very definitive prediction of the magnitude and
sign of the ratio of the two ratios,

\begin{equation}
\frac{g_{\pi NN^{\ast}(1440)}}{g_{\pi NN}}=0.53 \; \frac{g_{\sigma
NN^{\ast}(1440)}}{g_{\sigma NN}} \, ,
\end{equation}

\noindent which is an exportable prediction of our model.

For the sake of completeness we give the values of $g_{\sigma NN^*(1440)}$
and $g_{\sigma NN}$, though one should realize that the corresponding form
factor ${\Lambda ^{2}}/(\Lambda^{2}-m_{\sigma }^{2})=2.97$ differs quite
much from 1. Moreover, the quark model dependence is quite strong what can
make nonsense any comparison to other values obtained in the literature
within a different framework. We get

\begin{equation}
\frac{g_{\sigma NN}^{2}}{4\pi } \frac{\Lambda ^{2}}{\Lambda
^{2}-m_{\sigma}^{2}}=72.4 \, ,
\end{equation}

\noindent i.e. $g_{\sigma NN}^{2}/{4\pi }=24.4$, and

\begin{equation}
\frac{g_{\sigma NN}}{\sqrt{4\pi }} \frac{g_{\sigma NN^{\ast}(1440)}}{\sqrt{%
4\pi }} \frac{\Lambda ^{2}}{\Lambda ^{2}-m_{\sigma }^{2}}=34.3 \, ,
\end{equation}

\noindent i.e. $g_{\sigma NN^{\ast}(1440)}^{2}/{4\pi }=5.5$.

Concerning the absolute value of $g_{\sigma NN^*}$ some caveats are in
order. Our value is scheme and quark-model dependent and should only be
sensibly compared with a value obtained in the same framework. As a matter
of fact, if we had extracted the quark model factor dependence from the
coupling constant ($e^{m_\sigma^2 b^2/2}$) \cite{JULI} the result
would have been $g_{\sigma NN^*(1440)}^{2} /4\pi=$1.14 that compares quite
well with the value given in Ref. \cite{OSET}, $g_{\sigma NN^*(1440)}^{2}/
4\pi =1.33$. With respect to the results
given in Ref. \cite{MADE} they are very
sensitive to both the decay width of the sigma meson into two pions and the
mass of the sigma as reflected in the large error bars given. 
Both quantities are highly undetermined in the Particle
Data Book \cite{PDGB}, the mass of the sigma 
being constrained between 400$-$1200 MeV
and the width between 600$-$1000 MeV. These values have been fixed
arbitrarily in Ref. \cite{MADE} to $m_{\sigma }=$500 MeV and $\Gamma
_{\sigma }=$250 MeV. Varying the mass of the sigma between 400 and 700 MeV
for a fixed width of 250 MeV, the coupling constant according to Eq. (9) of
Ref. \cite{MADE} varies between 0.18$-$2.54. Taking a width of 450 MeV the
resulting coupling is 0.27$-$1.64. In both cases, our value lies in the
interval given above what makes it compatible with the $N^*(1440)$ decay and
production phenomenology.

Let us finally mention that at short distances, the
interaction could be fitted in terms of two 
different Yukawa functions, one
depending on the meson mass, $m$, the other with a shorter
range depending on 
$\sqrt{(M_{N^(1440)}-M_{N}+m)m}$. These two Yukawa 
functions could be associated to the two diagrams
with different intermediate states ($mNN$ and $mNN^*(1440)$) appearing in
time ordered perturbation theory when an effective calculation at the baryonic
level is carried out (let us realize that in a quark calculation the
intermediate state is always $mqq$, the $N-N^*(1440)$ mass difference
being taken into account through the $N$ and $N^*(1440)$ wave functions).
For practical purposes, 
as done in previous works \cite{TRIT}, separable expansions
of the quark-based interactions can be performed and used in 
standard few-body calculations.

\section{Summary}

Starting from a quark-quark chiral symmetric interaction model and assuming
simple harmonic wave functions for $N$ and $N^{\ast }(1440)$ in terms of
quarks we have derived a transition $NN\rightarrow NN^{\ast }(1440)$
potential in an adiabatic approach. The Roper resonance has been taken as a
stable particle. Our results for $S$, $P$ and $D$ waves show significant
differences concerning the character of the interaction (attractive or
repulsive) at intermediate and longer distances with respect to the $%
NN\rightarrow NN$ and $NN^{\ast }(1440)\rightarrow NN^{\ast }(1440)$ cases
for the chosen $N^*(1440)$ overall phase. This has to do with the presence
of a node in the Roper wave function. On the contrary the short-range
interaction has the same character in all cases but the intensity gets
reduced in the $NN\rightarrow NN^{\ast }(1440)$ transition as a consequence
of the lesser similarity between initial and final states that makes the
Pauli principle to be less active. These results show that the usual
procedure of obtaining $NN^{\ast }$ interactions by a simple scaling of the $%
NN$ one should be handled with care.

The analysis of the asymptotic behavior of the potentials allows to 
determine the $\pi NN^{\ast }(1440)$ and $\sigma NN^{\ast }(1440)$ coupling
constants on the same foot than $\pi NN$ and $\sigma NN$ couplings. Ratios
between coupling constants of the type $\frac{g_{\pi NN^{\ast }(1440)}}{%
g_{\pi NN}}$ and $\frac{g_{\sigma NN^{\ast }(1440)}}{g_{\sigma NN}}$ are
obtained. These ratios, whose sign is ambiguous, are coupling scheme
independent and they have a softened quark model dependence (when compared
to the dependence of the value of each constant separately).
Furthermore the model allows the prediction of not only the magnitude but
also the relative sign between the two ratios.

We should finally notice that for dynamical applications our results should
be implemented by the inclusion of the $N^{\ast }(1440)$ width. Quantum
fluctuations of the two baryon center-of-mass, neglected here, could also
play some role. Though these improvements will have a quantitative effect we
do not think our predictions will be very much modified at a qualitative
level. In this sense they could serve either as a first step for more
refined calculations or as a possible guide for phenomenological
applications.

\acknowledgements
We would like to thank to E. Oset for a careful reading 
of the manuscript and useful comments and discussion. B.J.
thanks Ministerio de Ciencia y Tecnolog\'{\i}a for 
financial support. A.V.
thanks Ministerio de Educaci\'{o}n, Cultura y Deporte of Spain for
financial support through the Salvador de Madariaga program. This work has
been partially funded by Direcci\'{o}n General de Investigaci\'{o}n Cient%
\'{\i}fica y T\'{e}cnica (DGICYT) under Contract No. BFM2001-3563, by Junta de
Castilla y Le\'{o}n under Contract No. SA-109/01, and by EC-RTN, Network
ESOP, contract HPRN-CT-2000-00130.

\begin{table}[tbp]
\caption{ Quark-model parameters.}
\label{table1}
\begin{tabular}{cccc}
& $m_q ({\rm MeV})$ & 313 &  \\ 
& $b ({\rm fm})$ & 0.518 &  \\ 
\tableline & $\alpha_s$ & 0.485 &  \\ 
& $\alpha_{ch}$ & 0.027 &  \\ 
& $m_\sigma ({\rm fm^{-1}})$ & 3.42 &  \\ 
& $m_\pi ({\rm fm^{-1}})$ & 0.70 &  \\ 
& $\Lambda ({\rm fm^{-1}})$ & 4.2 & 
\end{tabular}
\end{table}

\begin{figure}[tbp]
\caption{Different diagrams contributing to the $NN \to NN^*(1440)$
interaction. The wavy line denotes an excited quark on the $1s$ shell and
the dashed line stands for an excited quark on the $0p$ shell. We have
labeled the diagrams attending to their topological equivalence, although
they involve interactions between excited or non excited quarks. This
simplified notation will be used in the next figures to separate the
different contributions to the interaction.}
\label{fig1}
\end{figure}

\begin{figure}[tbp]
\caption{$NN \to NN^*(1440)$ potential for (a) the $^1S_0$ partial wave, (b)
the $^3S_1$ partial wave, and (c) the long-range part of the $^1S_0$ partial
wave. We have denoted by the long-dashed, dashed, dotted, and dot-dashed
lines, the central OPE, OSE, OGE, and the tensor contributions,
respectively. By the solid line we plot the total potential.}
\label{1s0de}
\end{figure}

\begin{figure}[tbp]
\caption{Same as Fig. \ref{1s0de} but for (a) the $^1P_1$ partial wave, (b)
the $^3P_0$ partial wave, and (c) the long-range part of the $^1P_1$ partial
wave.}
\label{1p1de}
\end{figure}

\begin{figure}[tbp]
\caption{Same as Fig. \ref{1s0de} but for (a) the $^1D_2$ partial wave and
(b) the $^3D_1$ partial wave.}
\label{1d2de}
\end{figure}

\begin{figure}[tbp]
\caption{$NN \to NN^*(1440)$ potential for (a) the $^1S_0$ partial wave and
(b) the $^1P_1$ partial wave. We have made explicit the contribution of the
different diagrams shown in Fig. \ref{fig1}, with the convention explained
in the caption.}
\label{3s1di}
\end{figure}

\begin{figure}[tbp]
\caption{(a) Asymptotic behavior of the one-pion exchange $^1S_0$ $NN \to
NN^*(1440)$ potential (solid line). The dashed line denotes the fitted curve
according to Eq. (\ref{lrg}). (b) Same as (a) but for the one-pion exchange $%
^1S_0$ $NN \to NN$ potential.}
\label{nr1s0lrg}
\end{figure}

\begin{figure}[tbp]
\caption{(a) Asymptotic behavior of the one-sigma exchange $^1S_0$ $NN \to
NN^*(1440)$ potential (solid line). The dashed line denotes the fitted curve
according to Eq. (\ref{slrg}). (b) Same as (a) but for the one-sigma
exchange $^1S_0$ $NN \to NN$ potential.}
\label{nr1s0silrg}
\end{figure}

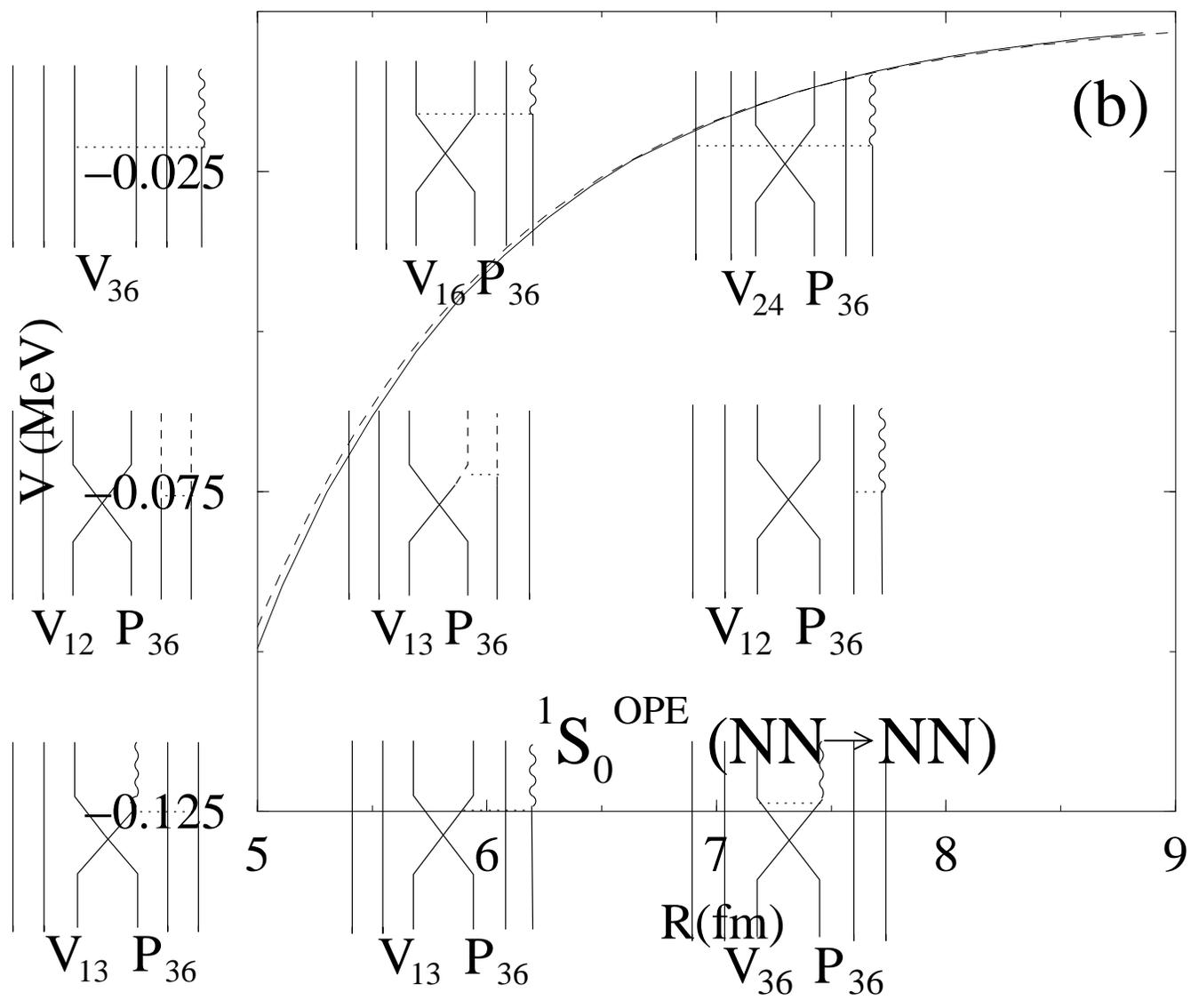
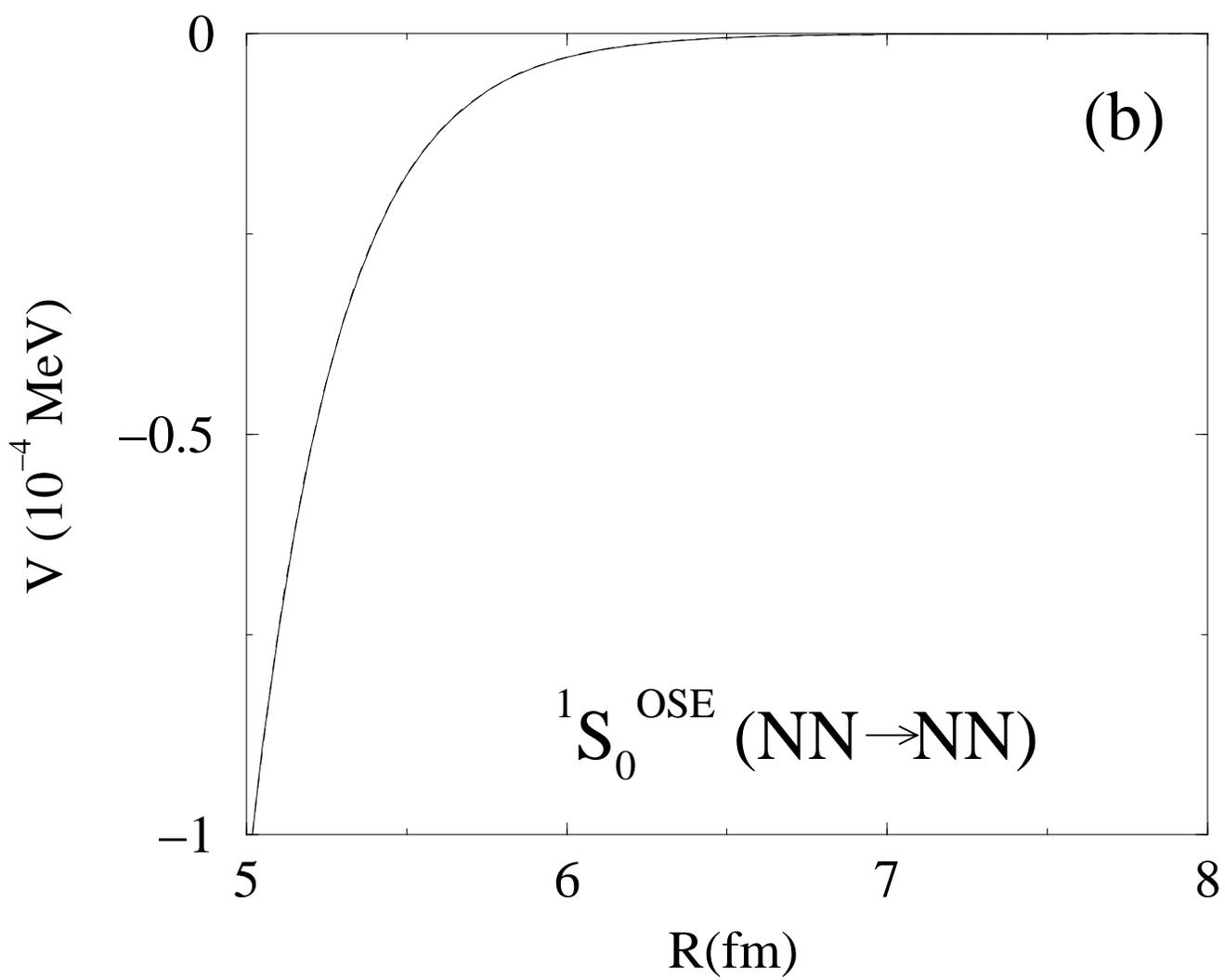
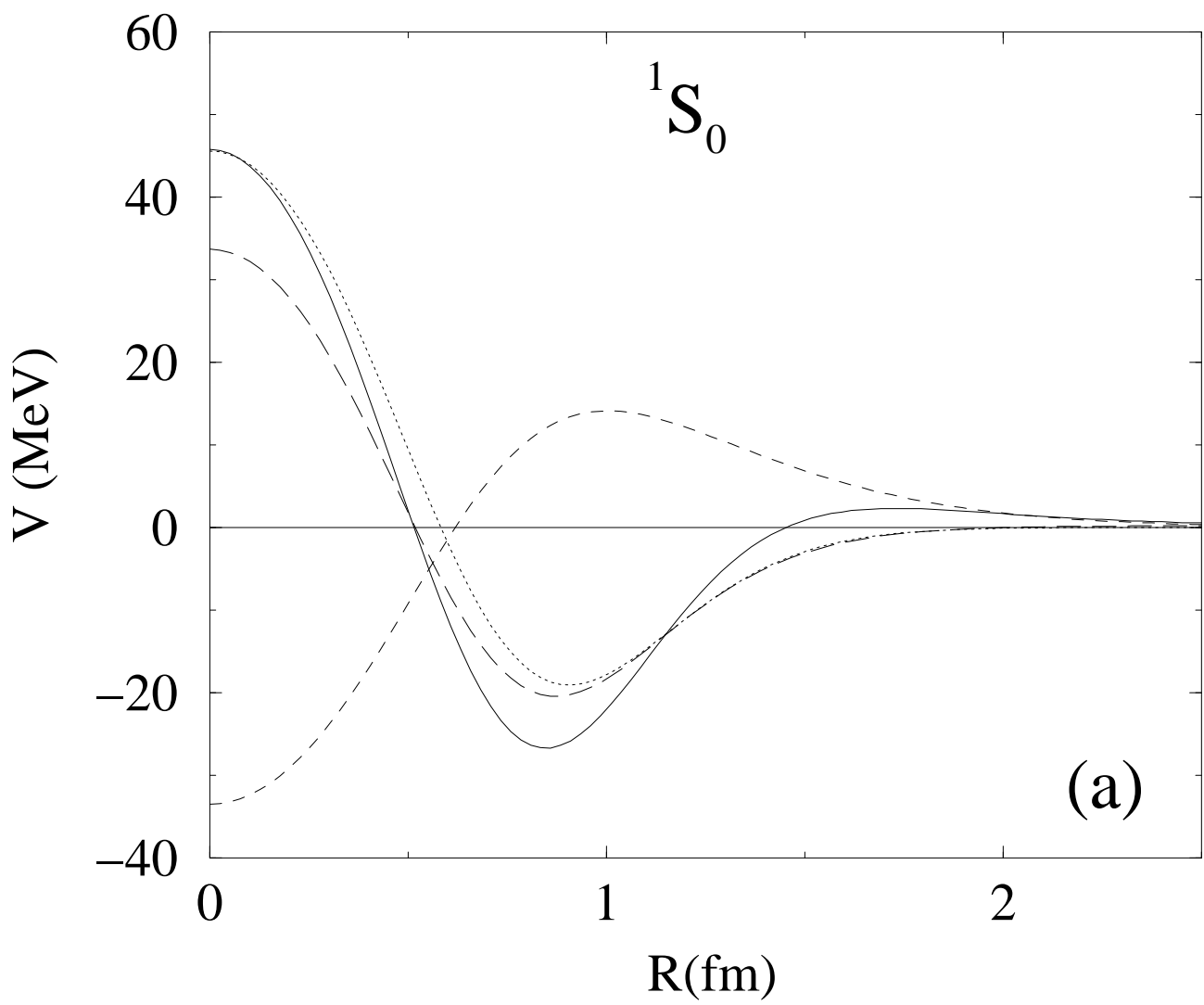
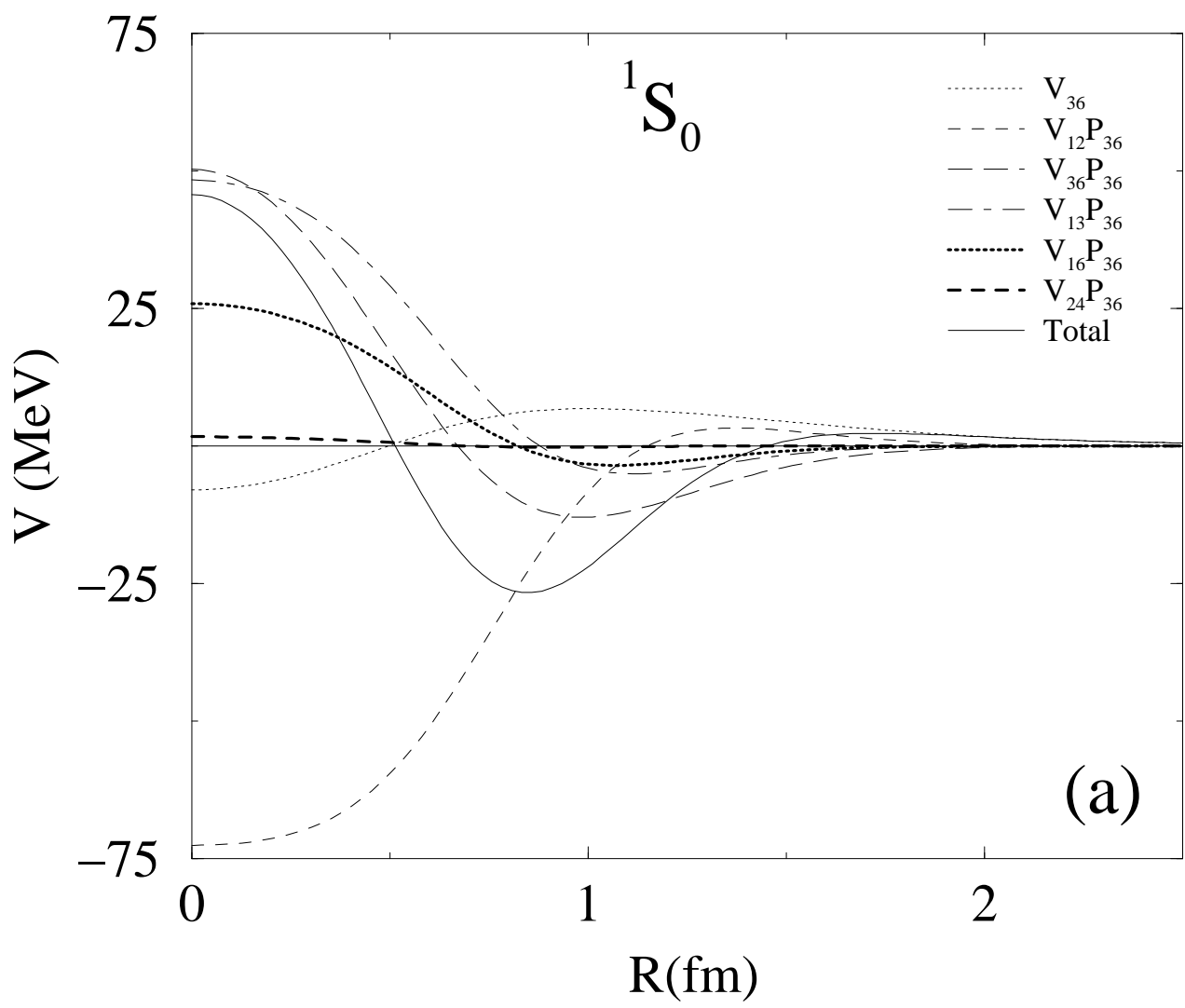
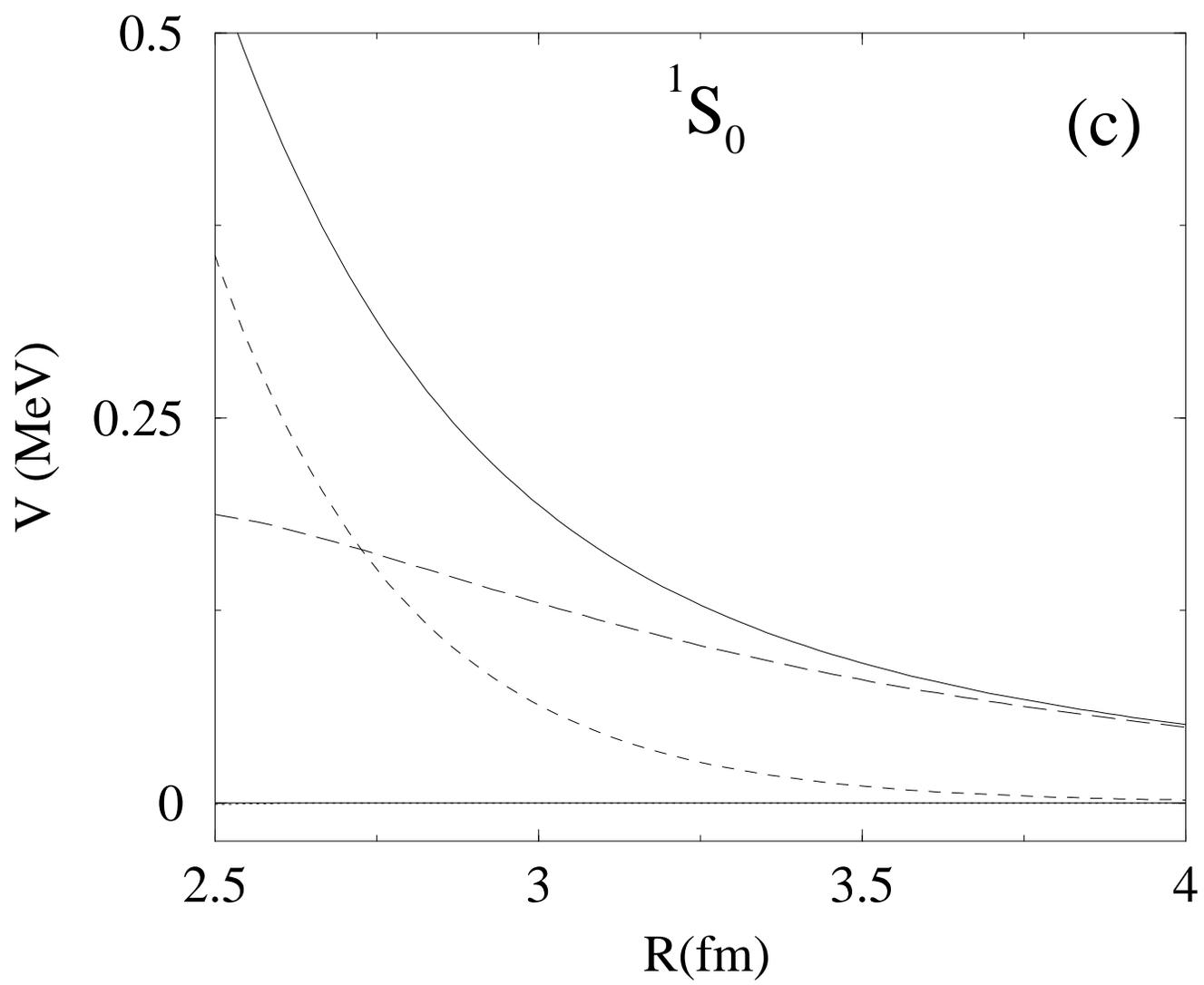
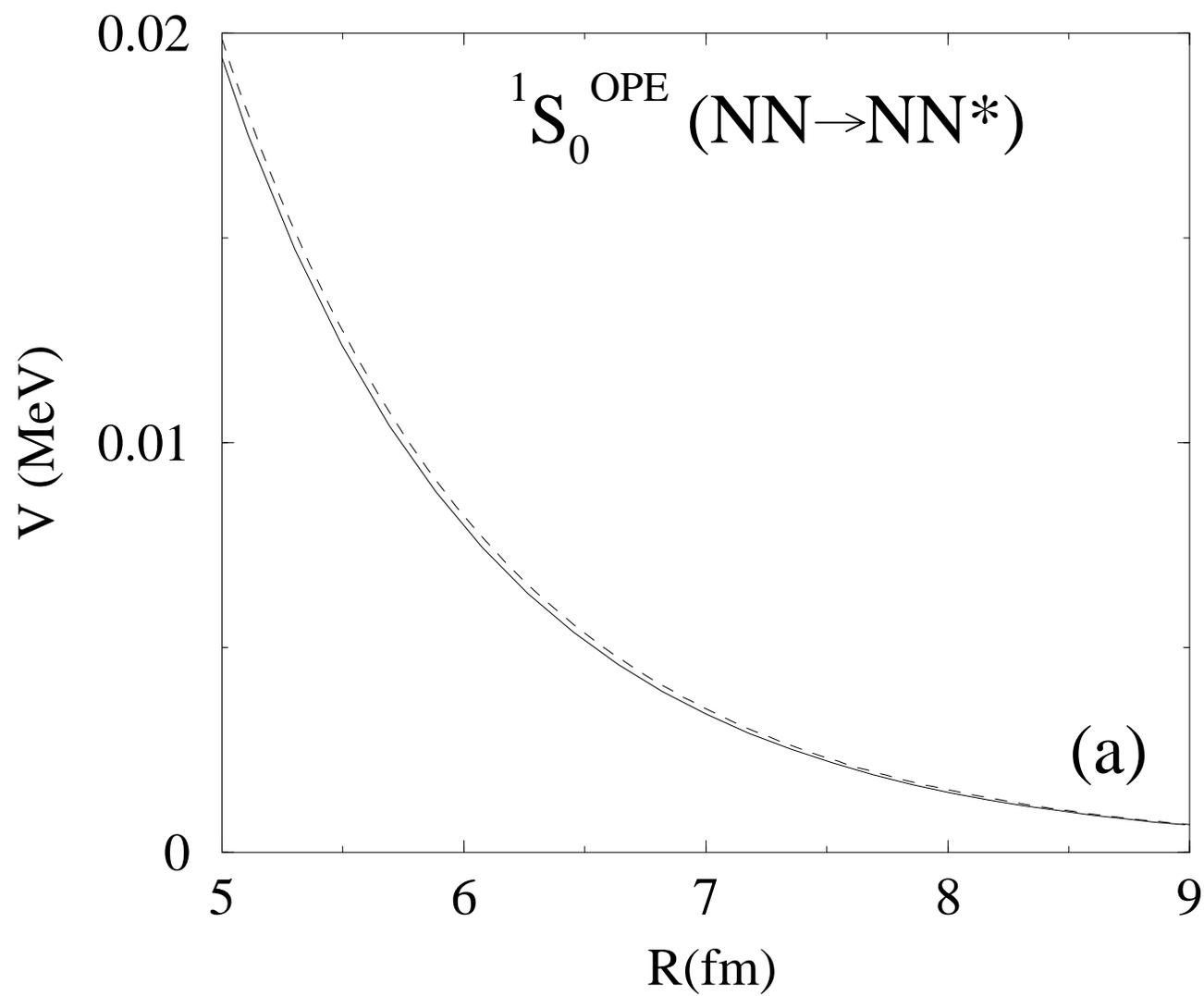
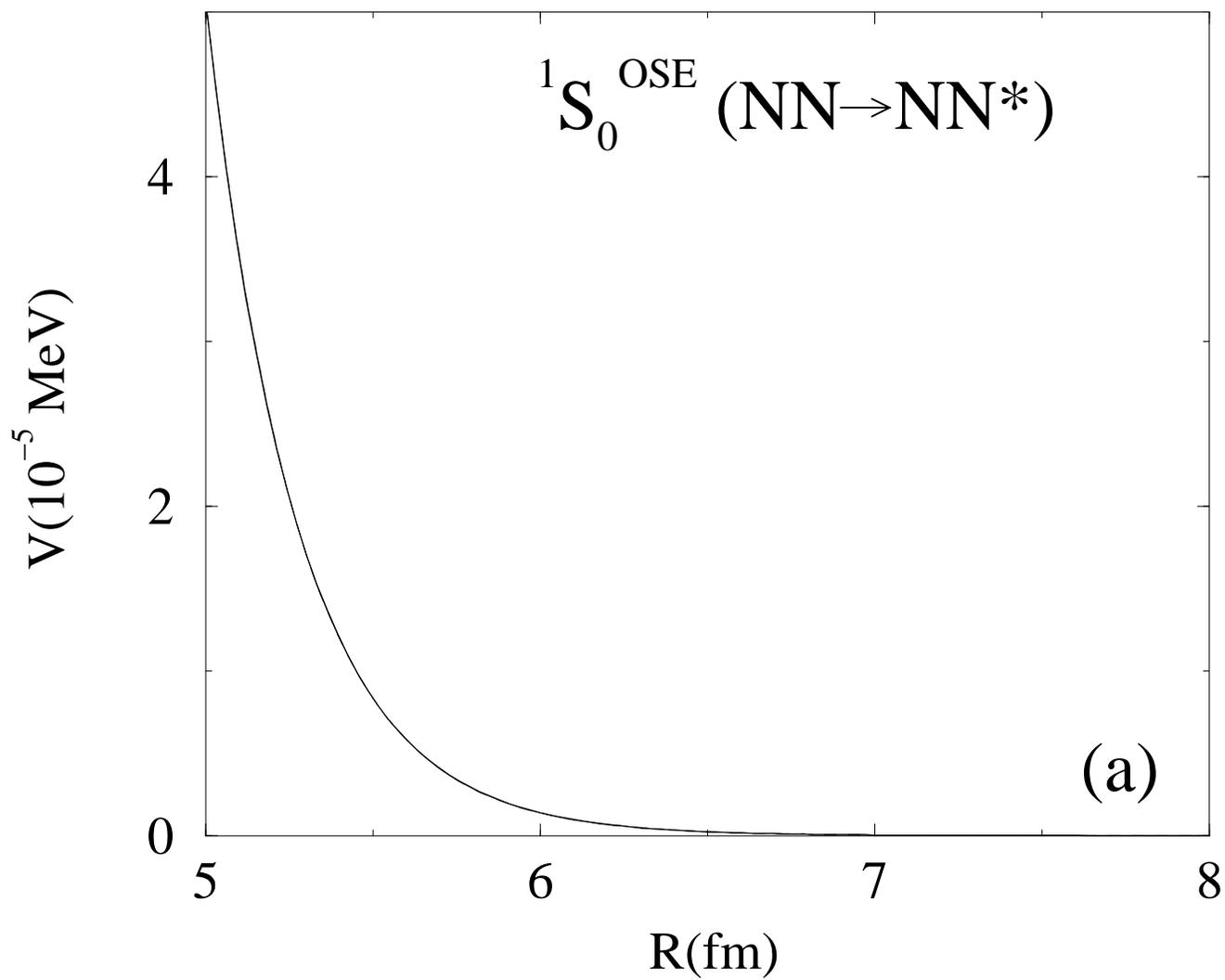
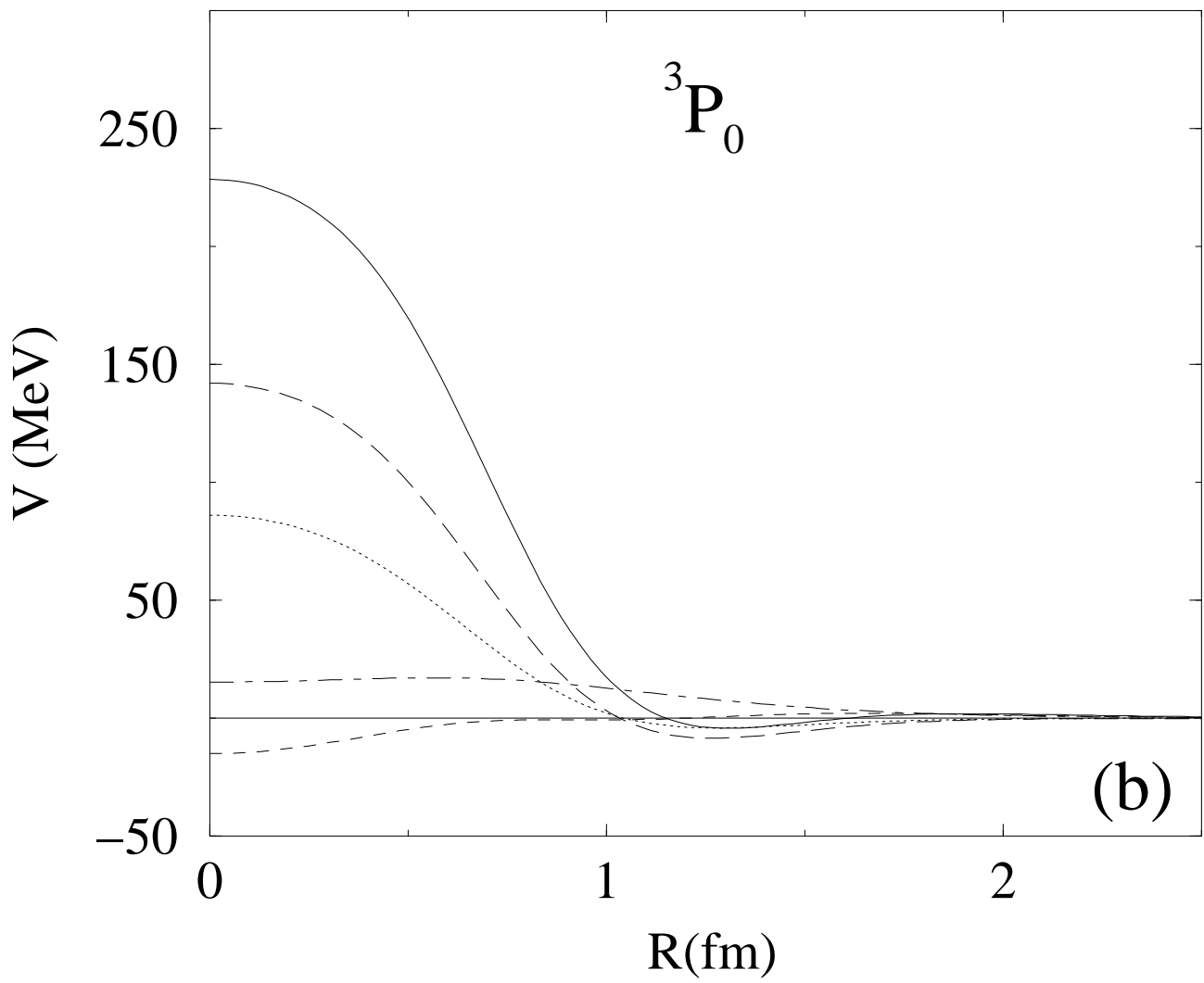
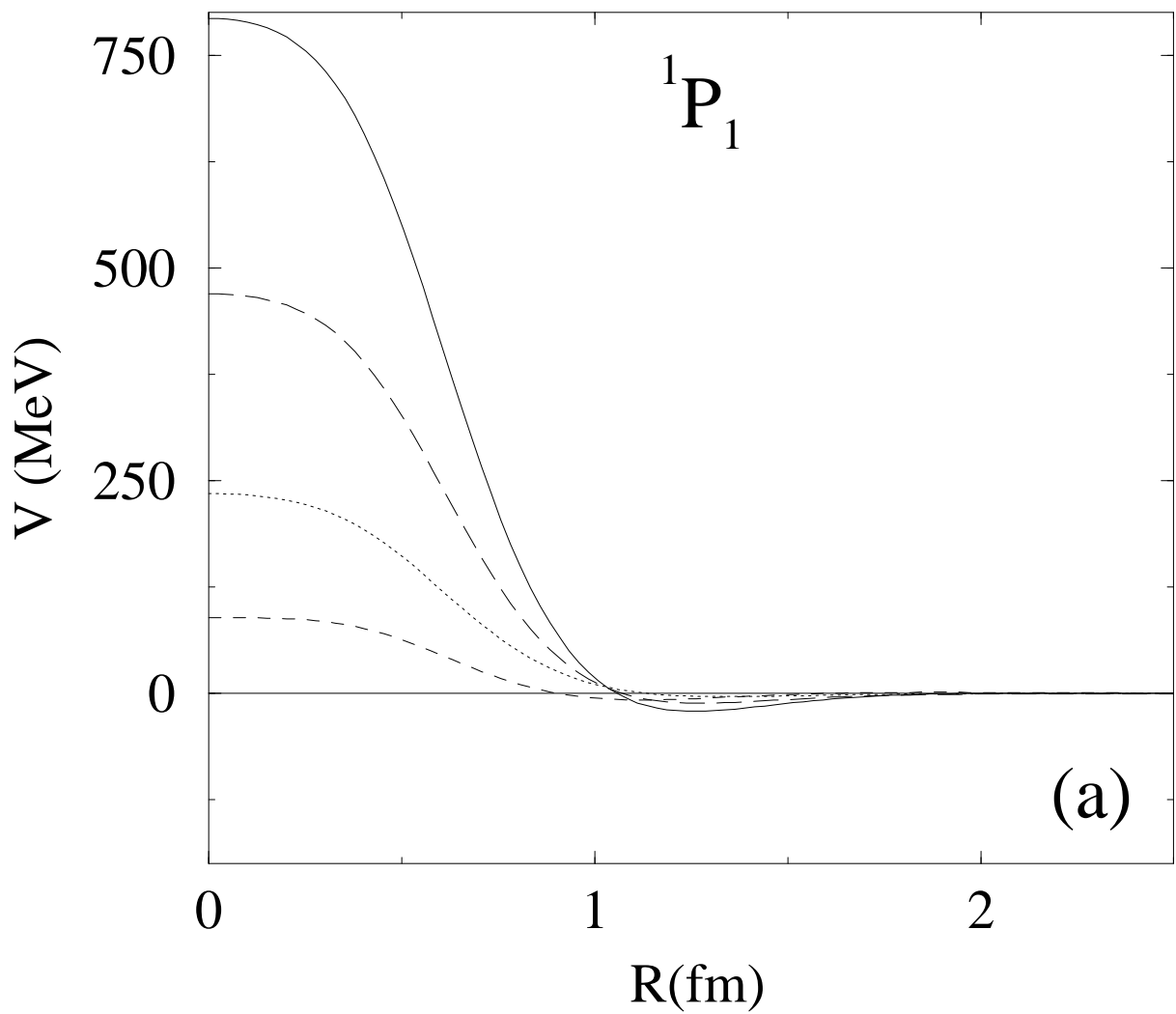
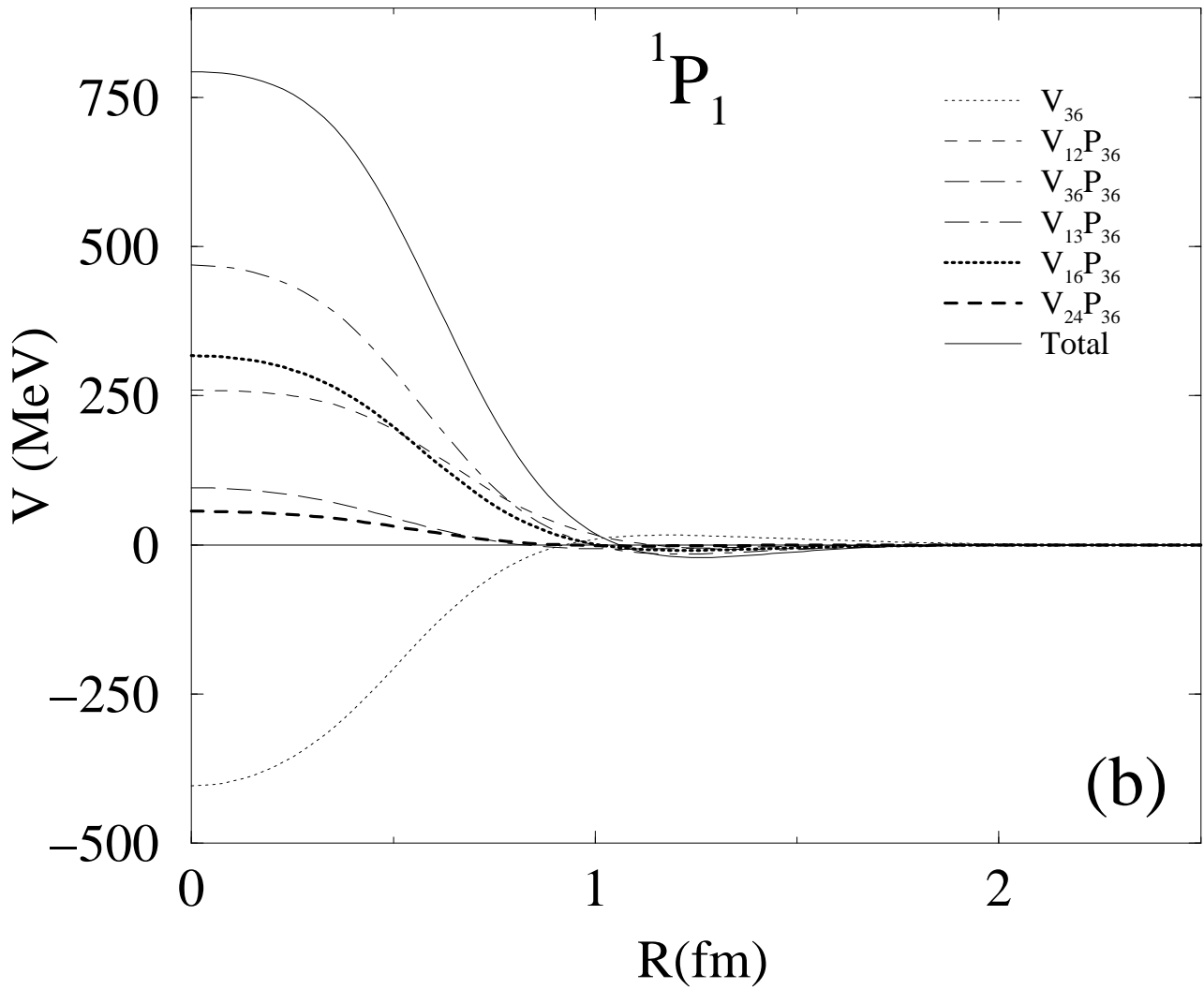
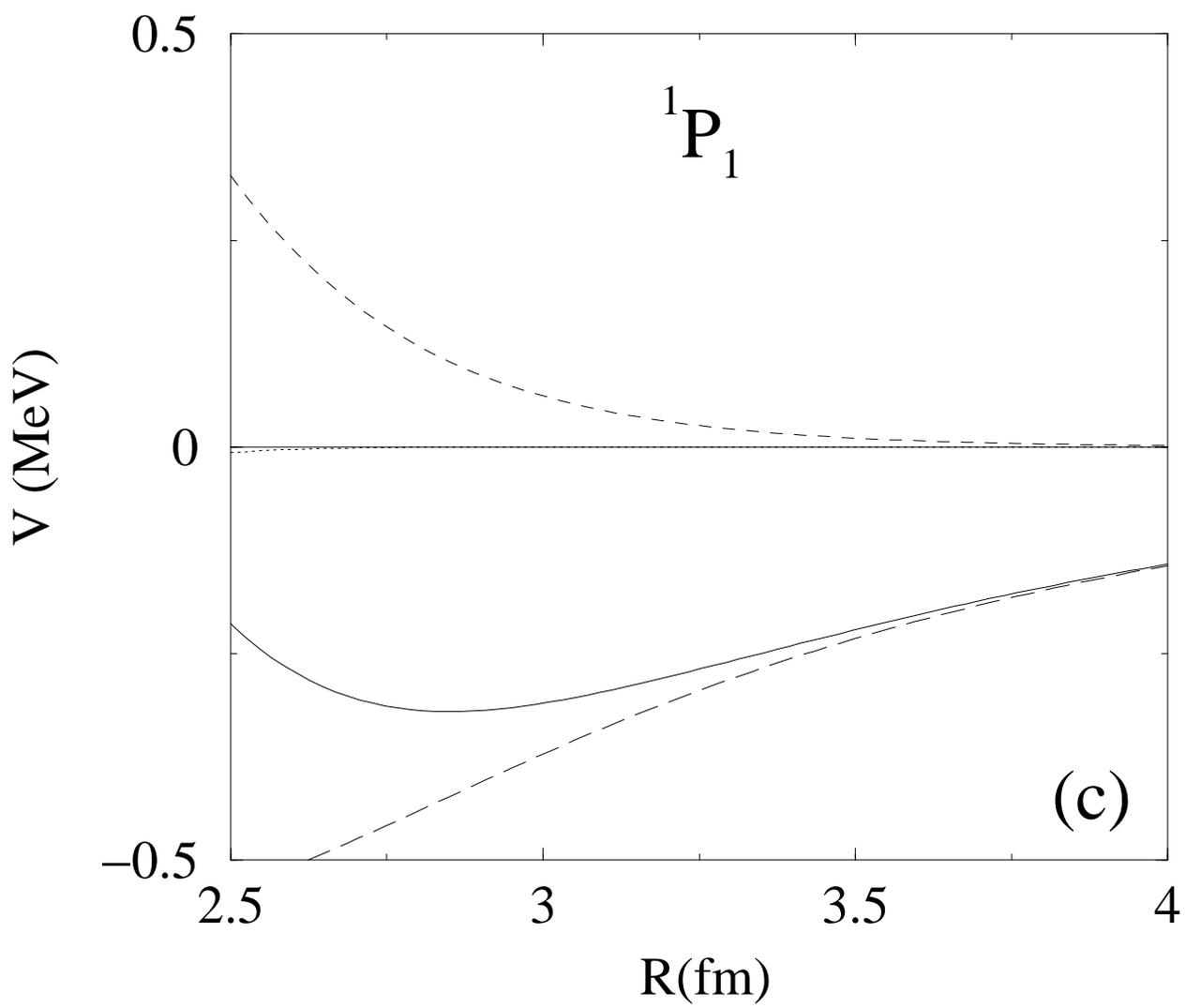
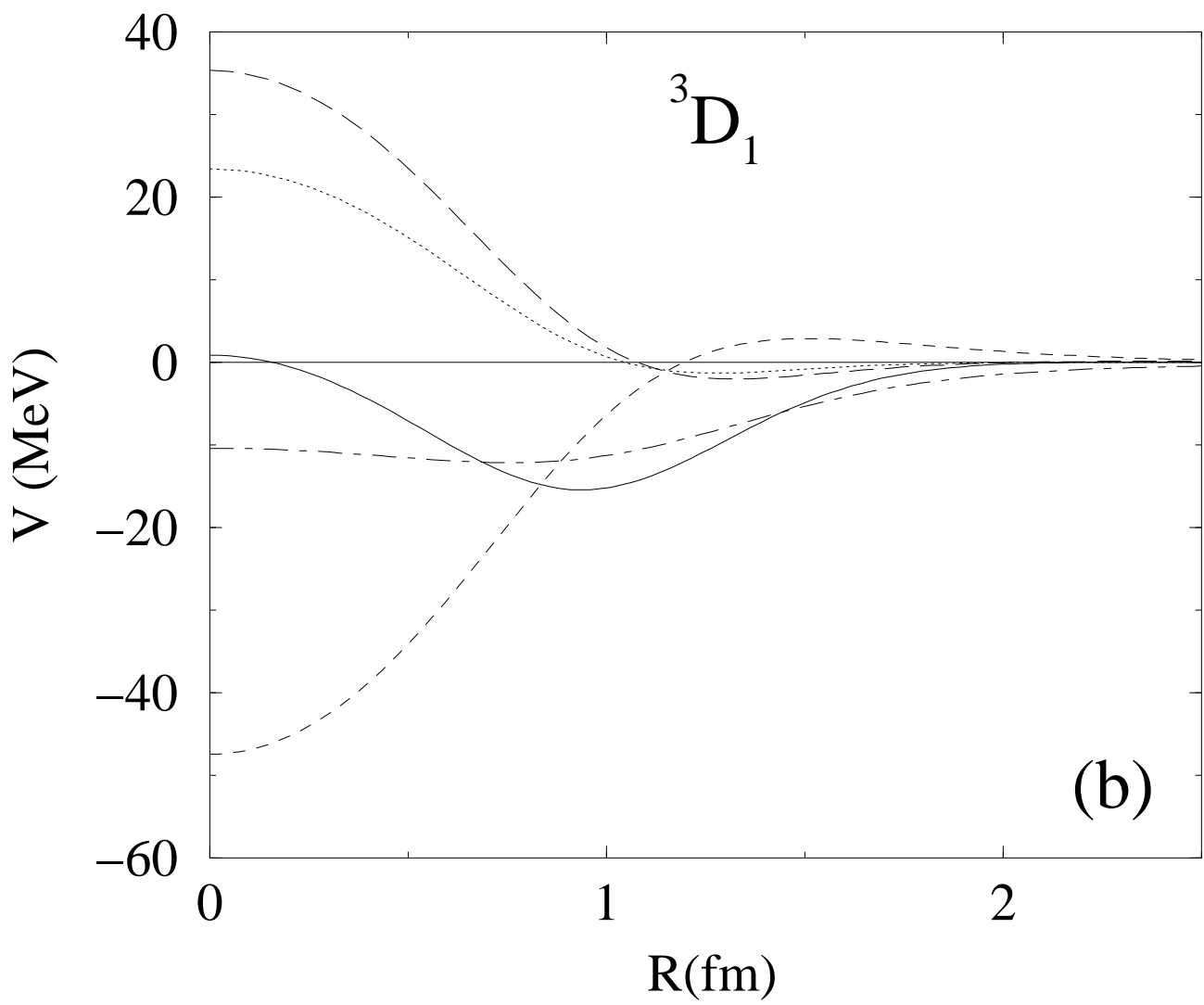
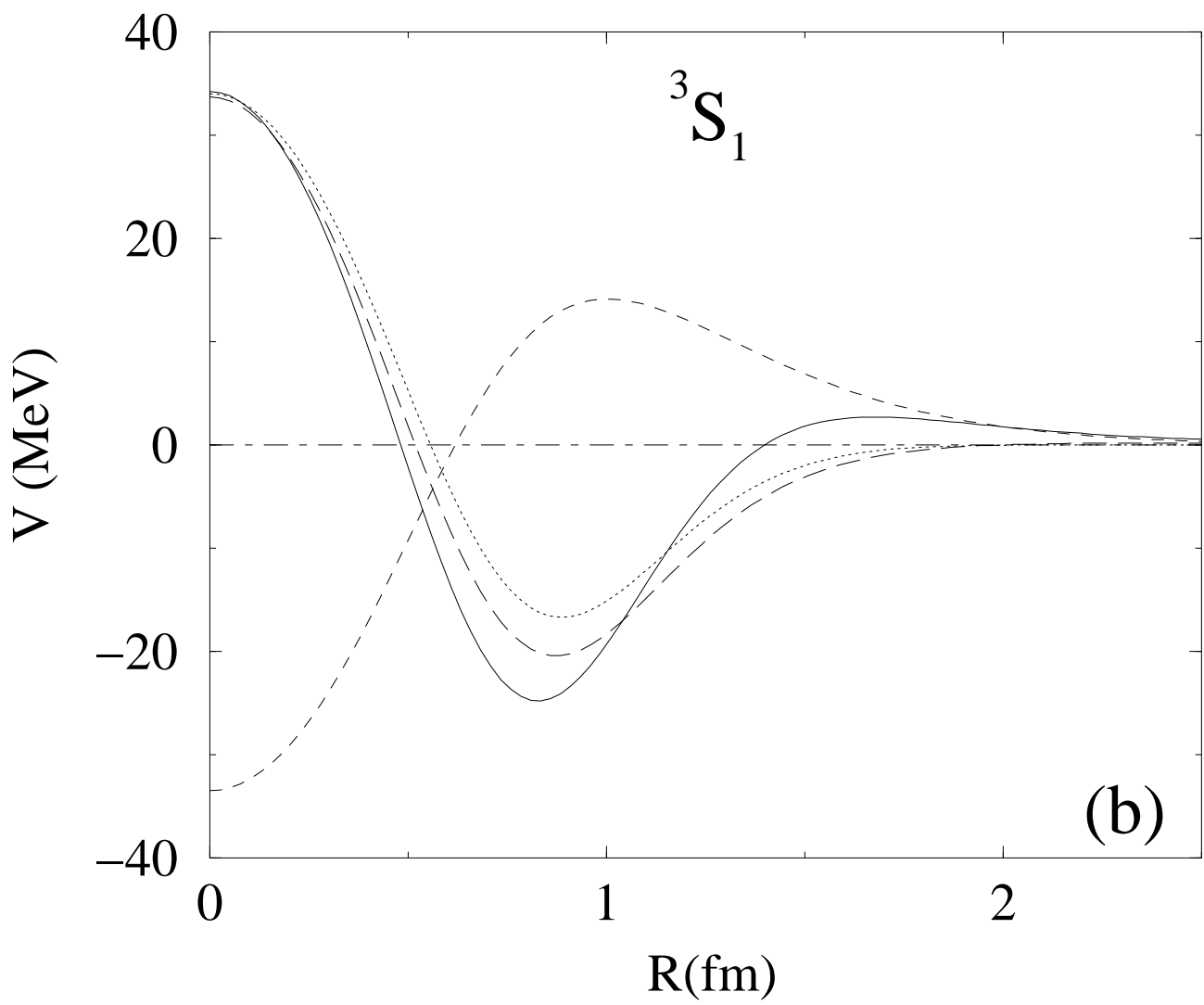
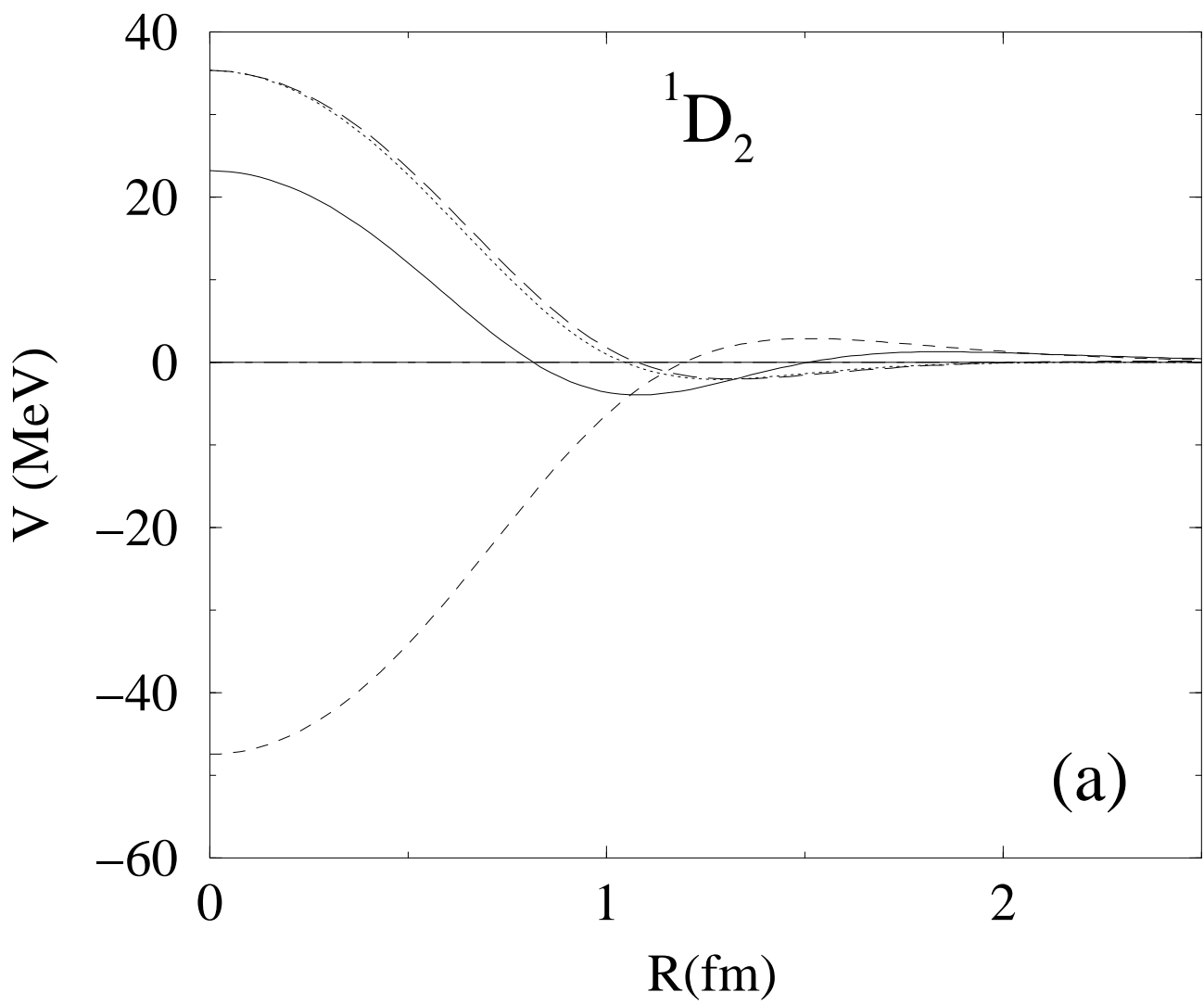
\end{document}